\documentclass[12pt]{article}

\usepackage{graphics} 
\usepackage{epsfig} 
\usepackage{mathptmx} 
\usepackage{amsmath} 
\usepackage{amssymb}  
\usepackage{color}
\usepackage{float}
\newcommand{\norm}[1]{\left\lVert#1\right\rVert}
\title{\LARGE \bf
Parsimonious  Volterra System Identification
}

\author{Sarah Hojjatinia$^{1}$, Korkut Bekiroglu$^{2}$, Constantino M. Lagoa$^{3}$
\thanks{*This work  is partially funded  by the National Science Foundation (NSF) Grant CNS-1329422.}
\thanks{$^{1}$Sarah Hojjatinia is with the School of Electrical Engineering and Computer Science,
        The Pennsylvania State University, University Park, PA, USA
        {\tt\small szh199@psu.edu}}%
\thanks{$^{2}$Korkut Bekiroglu is with the School of Electrical and Electronic
	Engineering, Nanyang Technological University, 639798 Singapore,
	{\tt\small kbekiroglu@ntu.edu.sg}}
    \thanks{$^{3}$Constantino M. Lagoa is with Faculty of the School of Electrical Engineering and Computer Science,
    	The Pennsylvania State University, University Park, PA, USA
    	{\tt\small lagoa@psu.edu}}%
}

\date{}
\newtheorem{Remark}{Remark}

\begin{document}

\maketitle
\thispagestyle{empty}
\pagestyle{empty}

\begin{abstract}

In this short paper, we aim at developing algorithms for  sparse Volterra system identification when the system to be identified has infinite impulse response. 
Assuming that the impulse response is represented as a sum of exponentials and
given input-output data, the problem of interest is to find the "simplest" nonlinear Volterra model which is compatible with the a priori information and the collected data. 
By simplest, we mean the model whose impulse response has the least number of exponentials.
The algorithms provided are able to handle both fragmented data  and measurement noise. Academic examples at the end of paper show the efficacy of proposed approach.

\end{abstract}

\section{INTRODUCTION}
The linear time invariant system identification problem has been widely addressed in the literature. There are two approaches for linear time invariant system identification; set membership identification algorithms  \cite{sanchez1998robust, sznaier2003control} and subspace identification algorithms \cite{van1993subspace, van1994n4sid, van2012subspace}.
Most techniques in parameter estimation of nonlinear systems, are dependent to linearizing the system first, and then using available toolboxes for linear system identification. When the data is noisy the ability to solve a nonlinear system identification becomes more difficult, as linearizing based approaches fail. Therefore, finding a method which is precise and computationally effective is of great value.

Approximating nonlinear systems with parsimonious models has the potential to be precise and computationally efficient. Sparsity in the identification of linear and nonlinear systems has been the focus of recent research. 
 Bekiroglu \textit{et al.} \cite{bekiroglu2014parsimonious}, Shah \textit{et al.}\cite{shah2012linear}, and Yilmaz \textit{et al.} \cite{yilmaz2013efficient} determined low order models from noisy input-output data for linear systems, using the concepts of sparse matrix and signal recovery. The authors showed how these approaches can be applied for the case of scattered data and multiple runs of the same system with different initial conditions, for single input single output linear systems  \cite{bekiroglu2014parsimonious, shah2012linear, yilmaz2013efficient}. Bekiroglu \textit{et al.} \cite{bekiroglu2015low} extended the idea of low order linear system identification from noisy data, to the multiple input multiple output systems.
 
We extend the concept of sparse system identification to nonlinear systems. 
More precisely,  we consider systems in the form of a Volterra  series, which represents a large class of nonlinear systems.  
There is a large body of literature on the subject of Volterra models
and it mainly addresses
 nonlinear extensions of finite impulse response (FIR) models  \cite{ACS:ACS1272}. Their structures are similar, with the first order Volterra system being a linear FIR model  \cite{Nikolaou2000}.
 
Volterra models are linear with respect to kernel coefficients which aids in comparison to other nonlinear models. 
Similar to the approaches for linear FIR models,  a least squares approach can be applied to the  identification of Volterra models \cite{Nikolaou2000}.
 L. Yao, \textit{et al.} \cite{236898} used least square criteria and introduced recursive approximation and estimation algorithm for Volterra system identification, which is dependent on the number of non-zero Volterra kernel coefficients. 
 \subsection{Motivation}
 In contrast with previous work, our aim is to perform infinite impulse response identification from a  finite number of noisy measurements. 
 The objective is to identify the sparsest nonlinear Volterra system which is compatible with noisy data collected and a-priori information. Algorithms provided for identifying the  Volterra system aim at obtaining a compact description of the model for the infinite impulse response case. 
 
 
 \subsection{Paper Outline}
The paper is structured as follows:   
after this introduction, in Section \ref{pre} formulation and structure of second order Volterra system is defined. Section \ref{stat} includes the problem statement. In Section \ref{problem} the sparse identification problem  is formulated. Algorithms to tackle the problem are defined in Section \ref{algo}. Numerical results are shown in Section \ref{result}.
And finally Section \ref{conclusion} concludes the paper with some possible future work.

\section{VOLTERRA SYSTEM} \label{pre}
For simplicity of exposition, in this paper we are going to concentrate on second order Volterra systems. However, the approach provided can readily be generalized to higher order Volterra models.
\subsection{Discrete-time System Representation}
Considering $x$ and $y$ as the input and output of system respectively, the truncated Volterra system of order $2$ with $L$ sample memory is defined as
\begin{equation} \label{eq:volterra2}
	\begin{aligned}
		y(n)= &h_{0}+\sum_{k_{1}=0}^{L-1} h_{1}(k_{1}) x(n-k_{1})+ \\
		&\sum_{k_{2}=0}^{L-1} \sum_{k_{1}=0}^{L-1} h_{2}(k_{1},k_{2}) x(n-k_{1}) x(n-k_{2})
	\end{aligned}
\end{equation} 
where $h_{0}$ is a constant. $h_{1}$ and $h_{2}$  are Volterra kernel coefficients that can be considered as  impulse response and higher order system impulse response of the system.

\subsection{Latrix-Vector Formulation of Volterra System}
Based on \cite{mathews2000polynomial}, the first $N$ measurements of second order Volterra system can be represented in standard matrix-vector form as


\begin{equation} \label{eq2}
	\mathbf{y}=X\mathbf{h}
\end{equation}
where $\mathbf{y}$ is the vector of output measurements, i.e
\begin{equation*} \label{eq:y}
	\mathbf{y}=
	\begin{bmatrix}
		y(0)  & y(1) & y(N-1)
	\end{bmatrix}^{T},
\end{equation*}
$N$ is the total number of measurements, 
and
\begin{equation} \label{eq:X}
X=
\begin{bmatrix}
1 & \mathbf{x_{1}}^{T}(0)  &  \mathbf{x_{1}}^{T}(0)  \otimes  \mathbf{x_{1}}^{T}(0) \\
1  & \mathbf{x_{1}}^{T}(1)  & \mathbf{x_{1}}^{T}(1)  \otimes  \mathbf{x_{1}}^{T}(1)  \\
\vdots & \vdots &\vdots \\
1 & \mathbf{x_{1}}^{T}(N-1)  & \mathbf{x_{1}}^{T}(N-1)  \otimes  \mathbf{x_{1}}^{T}(N-1) 
\end{bmatrix}
\end{equation}
where
\begin{align*} \label{eq:x1}
\mathbf{x_{1}}^{T}(n)=
\begin{bmatrix}
x(n) & x(n-1)  &  \cdots & x(n-(L-1)) 
\end{bmatrix},
\\
\forall n=0,\cdots, N-1
\end{align*}
where $\otimes$ denotes the Kronecker product.

%
\begin{equation}  \label{eq:h}
	\mathbf{h}=
	\begin{bmatrix}
		h_{0} & \mathbf{h}_{1}^T &  \mathbf{h}_{2}^T
	\end{bmatrix}^T
\end{equation}
~~~~~~~~~~~~ $\mathbf{h}_{1}^{T}=[ h_{1}(0) ~~ h_{1}(1)  ~~\cdots ~~ h_{1}(L-1)]$
\begin{equation*}  
	\mathbf{h}_{2}= vec(H_{2})
\end{equation*}
where $vec(H_{2})$ refers to vectorized version of the matrix $H_{2}$.
\begin{equation} 
	H_{2}=
	\begin{bmatrix}
		h_{2}(0,0)  & h_{2}(0,1)  & \cdots & h_{2}(0,L-1) \\
		h_{2}(1,0)  & h_{2}(1,1)  & \cdots & h_{2}(1,L-1) \\
		\vdots & \vdots &\vdots &\vdots \\
		h_{2}(N-1,0)  & h_{2}(N-1,1)  & \cdots & h_{2}(N-1,L-1) 
	\end{bmatrix}
\end{equation}
and
 $\mathbf{y} \in R^{N \times 1}, ~ X \in R^{N \times (1+L+L^{2})},~ \mathbf{h} \in R^{(1+L+L^{2}) \times 1}$.
 \newline
 \section{PROBLEM STATEMENT}\label{stat}
 \subsection{Infinite Impulse Response Representation}
 Since our objective is to identify Volterra models with infinite impulse response, we represent their impulse response  as a sum of exponentials. 
 More precisely $h_{1}$ is  represented as
 \begin{equation} \label{h1}
 h_{1}(k_{1})=\sum_{i} c_{1i}~a_{1i}(k_{1}) +\bar{c}_{1i}~\bar{a}_{1i}(k_{1}) 
 \end{equation}
where $c_{1i}$ is a complex number,
$\bar{c}_{1i}$ is the complex conjugate of $c_{1i}$,
${a}_{1i}(k_{1})$ is the atom which is considered as the sum of exponentials of the form:
 \begin{equation} \label{exp1}
a_{1i}(k_{1})=\alpha_{i}~p_{i}^{k_{1}-1}
\end{equation}
where $\alpha_{i}$s are the scaling factors and $p_{i}$s are  ''poles'' inside the unit circle.
Also, $\bar{a}_{1i}(k_{1}) $ is the complex conjugate of $a_{1i}(k_{1})$.

In a similar way, 
 $h_{2}$ is represented as
\begin{equation} \label{h2}
\begin{aligned}
h_{2}(k_{1},k_{2})=\sum_{i} &c_{2i}~a_{2i}(k_{1},k_{2}) +\bar{c}_{2i}~\bar{a}_{2i}(k_{1},k_{2}) 
\end{aligned}
\end{equation}
where the atoms $a_{2i}(k_{1},k_{2})$s are of the form:
\begin{equation} \label{exp2}
a_{2i}(k_{1},k_{2})=\beta_{i}~p_{1i}^{k_{1}-1}~p_{2i}^{k_{2}-1}
\end{equation}
where $\beta_{i}$s are the scaling factors, $p_{1i}$ and $p_{2i}$ are the "poles", and $\bar{a}_{2i}(k_{1},k_{2})$ is the complex conjugate of $a_{2i}(k_{1},k_{2})$.

Our objective  is finding the second order Volterra model of the form of equation (\ref{eq:volterra2}) that uses the least possible number of exponentials.
We aim to have a simple representation of Volterra model which can be used for simulation, control design and other purposes.
More precisely we aim at solving the following problem.

 \subsection{Problem}
Given

\begin{itemize} 
\item  \textit{  A compact set $D$ contained in the unit circle }
\item  \textit{   An unknown Volterra system (\ref{eq2}), whose impulse response is a sum of exponentials whose ``poles'' are in~$D$.}
\item  \textit{   An input sequence $x(k)$, applied to system, for $k=1,2,\ldots,N$  }
\item  \textit{   A noisy output sequence $y_{\eta}(k)$, for $k=
1,2,\ldots,N$, given by $y_{\eta}(k)=y(k)+\eta(k)$ }
  \textit{   where $\eta(k)$ denotes a noise sequence bounded by a known constant, i.e. \mbox{$\norm{\eta(k)}_{2} \;\leq \eta_{max}$}  }
\end{itemize}	
\textit{find the most parsimonious Volterra model which explains the input-output
pair within a given bound on estimation error~($\epsilon$).}

Summarizing, we assume the exponentials in equations~(\ref{exp1}) and~(\ref{exp2}) have their ``poles'' $p_{i}, p_{1i}$ and $p_{2i}$ in a  a given compact set $D$ and we aim at minimizing the number of exponentials in the representation of Volterra model while simultaneously having a model whose response is close to the collected data.
\subsection{Atomic Norm}

Elements of atomic set  $\mathcal{A}$ are called atoms. The atomic norm of a model $g$ is defined as:
\begin{equation} \label{gauge2}
\begin{aligned}
\norm{ g }_{\mathcal{A}}=\text{inf}~\{&\sum_{a_1 \in \mathcal{A}(D)} |c_{a_1}|+
\sum_{a_2 \in \mathcal{A}(D)} |c_{a_2}|:\\& ~~ g=\sum_{a_1  \in \mathcal{A}(D)} c_{a_1}~a_1+\sum_{ a_2 \in \mathcal{A}(D)}c_{a_2}~a_2\}.
\end{aligned}
\end{equation}
where
$\mathcal{A}(D)$ is the set of atoms related to the compact set $D$. Every function with poles in $D$ can be approximated as  a linear combination of atoms in the set $\mathcal{A}(D)$.
\begin{equation}
\begin{aligned}
\centering
\mathcal{A}(D)=&\mathcal{A}_{0}(D) \cup \mathcal{A}_{1}(D) \cup \mathcal{A}_{2}(D)\\
\mathcal{A}_{0}(D)=&\{a_{0}=1,~a_{1i}(k_{1})=0,~  a_{2i}(k_{1},k_{2})=0  \} \\
\mathcal{A}_{1}(D)=&\{a_{0}=0,~a_{1i}(k_{1})=\alpha_{i}~p_{i}^{k-1},~ a_{2i}(k_{1},k_{2})=0~ :\\
& ~ \forall \, p_{i} \in D  \} \\
\mathcal{A}_{2}(D)=&\{a_{0}=0,~a_{1i}(k_{1})=0,~  a_{2i}(k_{1},k_{2})=\beta_{i}~p_{1i}^{k_{1}-1}~p_{2i}^{k_{2}-1}~: ~\\
&~\forall \,  p_{1i},\, p_{2i} \in D   \} \\
\end{aligned}
\end{equation}
\subsection{Higher Order Volterra Models}
In this part we give some insight toward  the generalization of proposed  approach to higher order Volterra models.  In general, the Volterra system  is  in the form of 
\begin{equation} \label{eq:3.1}
y(n)= h_{0}+\cdots+\sum_{k_{1}=0}^{\infty}\dots \sum_{k_{m}=0}^{\infty} h_{m}(k_{1},\dots,k_{m})\prod_{i=1}^{m} x(n-k_{i})+\dots
\end{equation}
where $h_{0}$ is a constant, and $h_{m}s,~ \forall m=1,\cdots,\infty$ are Volterra kernel coefficients.
Applicable  Volterra systems are truncated to some predefined order.
The truncated Volterra system of order $m$ with $L$ sample memory is 
defined as
\begin{equation} \label{eq:3.2}
y(n)= h_{0}+\cdots+\sum_{k_{1}=0}^{L-1}\dots \sum_{k_{m}=0}^{L-1} h_{m}(k_{1},\dots,k_{m})\prod_{i=1}^{m} x(n-k_{i})
\end{equation} 
%
where for $m=1$, the Volterra system is
a linear system. By considering $m \geq 2$ the Volterra system represents a nonlinear system. 
For Volterra system of order $m$, kernel coefficient $h_{m}(k_{1},\dots,k_{m})$ will be defined in the same way of equations \eqref{h2} and \eqref{exp2}.
\begin{equation} \label{h2p}
\begin{aligned}
h_{m}(k_{1},\dots,k_{m})=\sum_{i} &c_{mi}~a_{mi}((k_{1},\dots,k_{m}) +\bar{c}_{mi}~\bar{a}_{mi}((k_{1},\dots,k_{m}) 
\end{aligned}
\end{equation}
where the atoms $a_{mi}((k_{1},\dots,k_{m})$s are of the form:
\begin{equation} \label{exp2p}
a_{mi}((k_{1},\dots,k_{m})=\zeta_{i}~P_{1i}^{k_{1}-1}~P_{2i}^{k_{2}-1}
\cdots~P_{mi}^{k_{m}-1}
\end{equation}
where $\zeta_{i}$s are the scaling factors, $P_{1i}$, $P_{2i}$, $\cdots$, and $P_{mi}$ are the poles. 
 Therefore, in a similar way we can formulate problem for higher order Volterra systems and apply the approach in this paper to identify the parsimonious model.

\section{FORMULATION OF THE SPARSE IDENTIFICATION PROBLEM} \label{problem}
As mentioned in the Introduction, in the identification of systems, whether linear or nonlinear, the concept of ``parsimonious representation'' plays an important role.
%
Kekatos and  Giannakis \cite{5997322} introduced compressed sampling approaches, in parsimonious identification of  Volterra and polynomial models.
Yilmaz \textit{et al.} \cite{7970196} introduced an algorithm for parsimonious linear system identification; specifically, it is a randomized algorithm which minimize a convex function over an atomic set; atomic set is considered to be scaled ball. 
In this paper we extend the idea of sparse  system identification for the case of nonlinear systems.

The problem of interest of this paper, parsimonious  identification of Volterra models with infinite impulse response, can be  formulated as
\begin{equation} \label{eq:optt}
\begin{aligned}
\min_{c} ~ & \text{cardinality} ~ \{c \neq 0\} \\
\text{s.t.} ~& \norm{  \mathbf{y}_{\eta}-X\mathbf{h} }_{2}^{2} \, \leq \epsilon\\
 &~ \text{equations}~ (\ref{eq:h}),~ (\ref{h1}), ~ \text{and}~ (\ref{h2}) 
\end{aligned}
\end{equation}
where $c$ is a vector containing all  coefficients, and
$\mathbf{y}_{\eta}$ is the vector of noisy measurements, i.e. 
\begin{equation*}
\mathbf{y}_{\eta}=\mathbf{y}+\eta
\end{equation*}
and $\mathbf{\eta}$ is noise 
\begin{equation*} \label{eq:n}
	\boldmath{\eta}=
	\begin{bmatrix}
		\eta(0)  & \eta(1) & \eta(N-1)
	\end{bmatrix}^{T}.
\end{equation*}
Also, the matrix $\mathbf{X}$, as in equation (\ref{eq:X}), is a structured matrix  containing input values, and $\mathbf{h}$ is a vector containing the first few entries of the impulse response.
Again, in the proposed approach,  $\mathbf{h}$ is assumed to be representable by a sum of exponentials. More precisely,
it has the form given by equations (\ref{eq:h}), (\ref{h1}), and (\ref{h2}).
%
$c_{i}$s are complex coefficients.

\begin{Remark}
	This optimization problem  can be readily modified to handle missing measurements.
The first constraint in equation (\ref{eq:optt}) can be replaced by $\norm{  \mathbf{y}_{\eta}-X\mathbf{h} }_{2}^{2} \, \leq \epsilon$ at the times when we have measurements.
\end{Remark} 

Minimizing cardinality subject to constraints is an NP-hard problem. It makes us to think about the relaxation of  original optimization problem in equation (\ref{eq:optt}) and using atomic norm.

\section{ALGORITHMS TO SOLVE THE PROBLEM}\label{algo}
\begin{figure} [h]
	{\includegraphics[width=1\columnwidth]{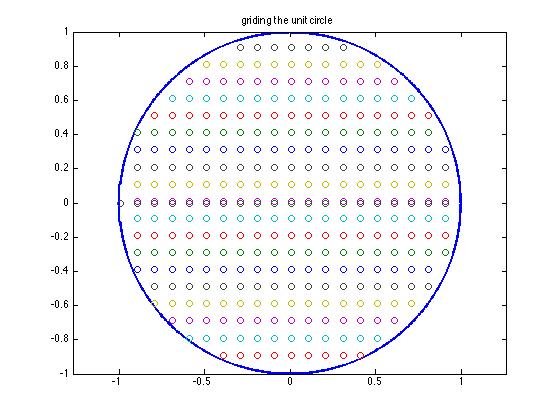} }
	\caption{A sample of griding the unit circle.} 
	\label{fig:grid}
\end{figure}

For solving or approximating the solution of the complex optimization problem in equation (\ref{eq:optt}), several algorithms are suggested in this section.
Therefore, the parsimonious Volterra system identification problem in (\ref{eq:optt}) can be approximated by:
\begin{enumerate}
\item Grid the set of atoms. One can do a griding of the set of poles as depicted in Figure \ref{fig:grid}   for  poles $p_{i}, p_{1i}$ and $p_{2i}$. Then, solve the mixed integer optimization problem 
\begin{equation} \label{eq:mip}
\begin{aligned}
\min~ &{\sum_{i}\sigma_{i}}  \\
\text{s.t.} ~& {\sigma_{i}}\in \{0,1\}\\
& |{c_{i}}|      \le M\sigma_{i} \\
 & ~ \text{equations}~ (\ref{eq:h}),~ (\ref{h1}), ~ \text{and}~ (\ref{h2})
\end{aligned}
\end{equation}
where $c_i$ is the coefficient associated with the exponential whose ``pole" is the $i$-th element of the grid and  $M$ is an upper bound on the maximum value of $ |{c_{i}}| $. This algorithm is again NP-hard.  Hence, two $l_{1}$ relaxations of the algorithm are introduced as follows:
%


\item Griding the set of atoms as above and using $l_{1}$ norm relaxation  of cardinality to obtain
$\norm{ c }_{1} $ sparse, as a surrogate of  $\text{cardinality} ~ \{c: c\neq 0\}$
\begin{equation} \label{eq:l1}
\begin{aligned}
\min_{\mathbf{h}} ~ & \norm{  \mathbf{h} }_{\mathcal{A}}  \\
\text{s.t.} ~& \norm{  \mathbf{y}_{\eta}-X\mathbf{h} }_{2}^{2} \, \leq \epsilon\\
 &~ \text{equations}~ (\ref{eq:h}),~ (\ref{h1}), ~ \text{and}~ (\ref{h2})
\end{aligned}
\end{equation}

\item Again, using $l_{1}$ as a surrogate of cardinality, the randomized version of Frank-Wolfe algorithm that was developed in \cite{7970196}; can be readily adapted to deal with the problem in this paper.
More precisely, one can use a randomized Frank-Wolfe algorithm to solve the problem

\begin{equation} \label{eq:opt1}
\begin{aligned}
\min_{\mathbf{h}} ~ & \norm{  \mathbf{y}_{\eta}-X\mathbf{h} }_{2}^{2} \, \\
\text{s.t.} ~& \norm{  \mathbf{h} }_{\mathcal{A}} \, \leq \tau\\
 &~ \text{equations}~ (\ref{eq:h}),~ (\ref{h1}), ~ \text{and}~ (\ref{h2})
\end{aligned}
\end{equation}

This approach has much less computational complexity than the other two. However, a closed form representation of the impulse response will not be obtained.  As a result, an additional step needs to be performed to identify the exponentials that are present in the identified impulse responses.

\end{enumerate}

\section{NUMERICAL RESULTS} \label{result}
In the examples to follow,  we solve the problem of sparse Volterra system identification with the $l_{1}$ norm relaxation. More precisely, we use algorithm 2  in Section \ref{algo}. 
Note that in using $l_{1}$ norm relaxation, scalings of the atoms ($\alpha_{i}$ and $\beta_{i}$) play a significant role on how sparse the solution can be and further research is needed to determine the ``optimal'' scaling values.
%

In the numerical examples presented,  $N$ measurements are considered. Random measurement noise  is added to true outputs. The unit circle has been gridded uniformly and random poles were picked to check as candidates for the system.  CVX is used as the convex optimization toolbox \cite{cvx}.
In this paper we use  scaling factors $\alpha_{i}=1$ and $\beta_{i}=1$.

Example one relates to a second order Volterra system, which is generated at random as a sum of 4 exponentials.  Random poles and coefficients selected for this algorithm are shown in Table \ref{tab:table11}.
A random uniform measurement noise with the maximum value, approximately $11.2\%$  of the mean  absolute value of outputs, is added.  Number of measurements is considered to be  $N=100$ in this example.

\begin{table*}[ht]
	\caption{Random selection of poles and coefficients for example 1.}
	\label{tab:table11}
	\centering
	\begin{tabular}{|c|c|c|} 
		\hline
		\textbf{First order poles} & \textbf{First order coefficients} & \textbf{Second order poles 1}  \\
		$p_i~\forall i=1,\cdots,4$ & $c_{1i}~\forall i=1,\cdots,4$ & $p_{1i}~ \forall i=1,\cdots,2$   \\
		\hline \hline
		-0.1375 + j 0.2731 & 1.8969 + j 0.0618 &0.2270 + j 0.1086 \\
		-0.1210 + j 0.3591	&  0.2834 -j 0.6773 & 0.4978 - j 0.4239 \\
		0.7844 + j 0.0577	& 1.9874 - j 0.2800  & \\
		-0.8890 - j 0.2277	&0.2142 - j 0.0328&\\
		\hline \hline
	\end{tabular}
	\begin{tabular}{|c|c|} 
	\hline
	 \textbf{Second order poles 2}& \textbf{Second order coefficients}  \\
	 $p_{2i}~ \forall i=1,\cdots,2$ & $c_{2i}~\forall  i=1,\cdots,2$  \\
	\hline \hline
	0.3591 - j 0.6873&1.5449 - j 1.2369 \\
   -0.7062 + j 0.5511 &  1.7244 - j 0.9657\\
	&   \\
	&\\
	\hline \hline
\end{tabular}
\end{table*}

\begin{figure} 
	\centering
	{\includegraphics[width=0.8\columnwidth]{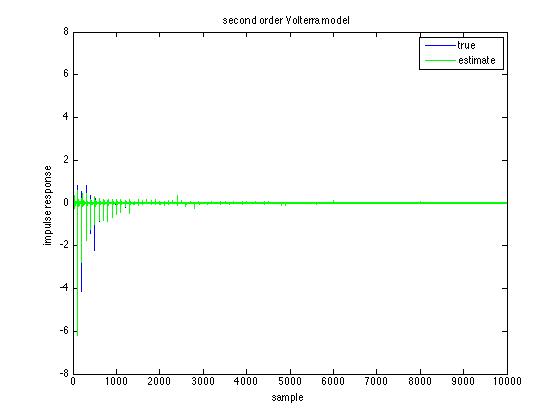} }
	\caption{True and estimated impulse response in second order Volterra system, example 1.} 
		\label{fig:order1-1}
\end{figure}

\begin{figure} 
	\centering
	{\includegraphics[width=0.8\columnwidth]{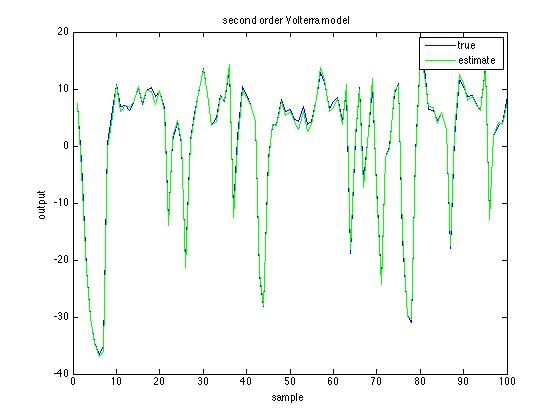} }
	\caption{True and estimated output in second order Volterra system, example 1.} 
	\label{fig:order1-2}
\end{figure}

\begin{figure} [h!]
	\centering
	\includegraphics[width=0.8\columnwidth]{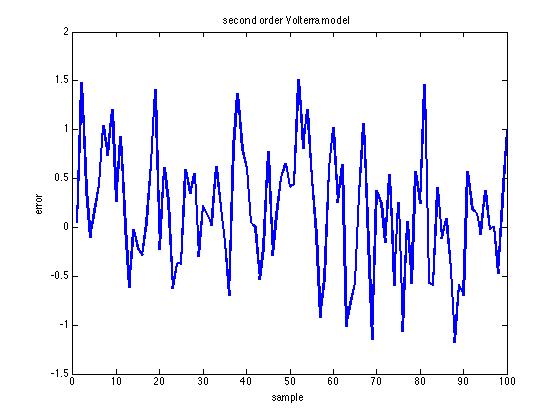} 
	\caption{Difference between true and estimated output in second order Volterra system, example 1.} 	
	\label{fig:order1-3}
\end{figure}

Impulse response of the estimated model ($\mathbf{h}_{2}$) compared to the true one is plotted in Figure~\ref{fig:order1-1}.  The true and estimated output are shown in Figure~\ref{fig:order1-2}; by true output we mean the output before adding noise. The difference between true and estimated output is shown in Figure \ref{fig:order1-3}. 
The estimated model is a sum of 5 exponentials, which shows a sparse identification of system. Therefore, having  optimal values for the scalings $\alpha_{i}$ and $\beta_{i}$, would result in  the true sparse system.

Example two relates to identification of a second order Volterra model with $N=150$ measurements.  Random poles and coefficients selected for this algorithm are shown in Table \ref{tab:table1}. Random  measurement noise with the maximum value, approximately $8\%$ of the mean  absolute value of outputs is added.
The corresponding poles and coefficients for this model are as Table \ref{tab:table1}.

\begin{table*}[ht]
		\caption{Random selection of poles and coefficients for example 2.}
	\label{tab:table1}
	\centering
	\begin{tabular}{|c|c|c|} 
	\hline
	\textbf{First order poles} & \textbf{First order coefficients} & \textbf{Second order poles 1}  \\
	$p_i~\forall i=1,\cdots,4$ & $c_{1i}~\forall i=1,\cdots,4$ & $p_{1i}~ \forall i=1,\cdots,2$   \\
	\hline \hline
	-0.6019 + j 0.2180 & -1.9283 - j 1.8763 & 0.3966 + j 0.6776 \\
	0.4813 - j 0.4971	&  -1.5213 - j 0.0245 & 0.0182 + j 0.0431 \\
	0.1924 - j 0.3459	&  1.8085 + j 1.4509 &\\
	0.8084 -j 0.0051	&1.9034 - j 1.0285&\\
	\hline \hline
\end{tabular}
\begin{tabular}{|c|c|} 
	\hline
	\textbf{Second order poles 2}& \textbf{Second order coefficients}  \\
	$p_{2i}~ \forall i=1,\cdots,2$ & $c_{2i}~\forall  i=1,\cdots,2$  \\
	\hline \hline
	-0.5943 - j 0.5600&0.5278 + j 0.2857 \\
	0.5027 + j 0.3444& -1.0269 + j 1.9269 \\
	&   \\
	&\\
	\hline \hline
	\end{tabular}
\end{table*}

\begin{figure} [h!]
	\centering
	{\includegraphics[width=0.8\columnwidth]{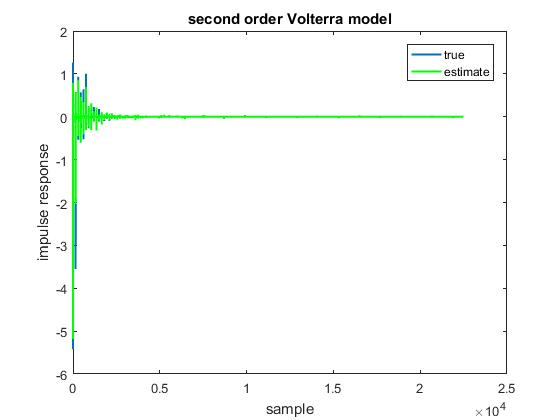} }
	\caption{True and estimated impulse response in second order Volterra system, example 2.} 
	\label{fig:order2-1}
\end{figure}

\begin{figure} 
	\centering
	{\includegraphics[width=0.8\columnwidth]{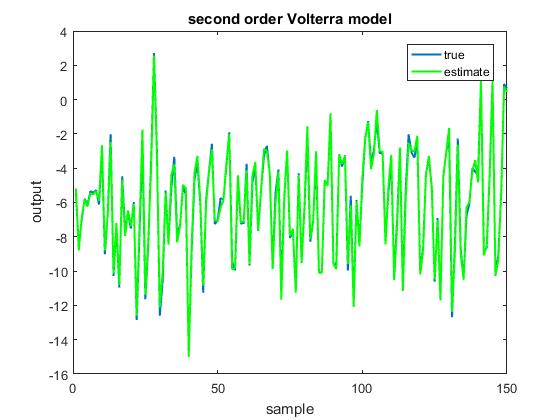} }
	\caption{True and estimated output in second order Volterra system, example 2.} 
	\label{fig:order2-2}
\end{figure}

\begin{figure} 	
	\centering
	\includegraphics[width=0.8\columnwidth]{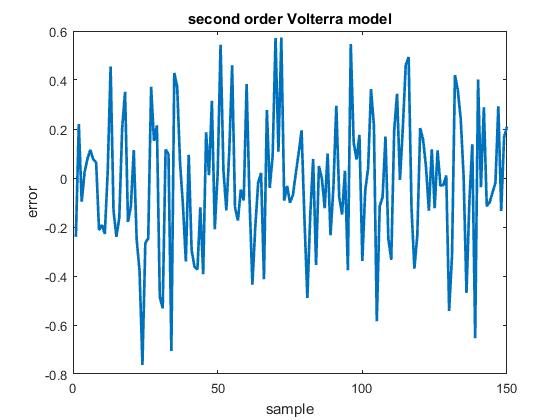} 
	\caption{Difference between true and estimated output in second order Volterra system, example 2.} 	
	\label{fig:order2-3}
\end{figure}

Figure \ref{fig:order2-1} shows the impulse response of the estimated model ($\mathbf{h}_{2}$)
compared to the true one. The true and estimated outputs of second order Volterra system are depicted in Figure \ref{fig:order2-2}. The difference between  estimated and true outputs is shown in Figure \ref{fig:order2-3}. 


As illustrated in the figures, the output of nonlinear system has been approximated efficiently. The impulse response shows good convergence. Coefficients of the identified Volterra model are close to the  coefficients of  true system. As a result, the sparsity has been achieved.

\section{CONCLUSION AND FUTURE WORK } \label{conclusion}
In this paper, we provide a novel approach to the problem of identifying  parsimonious models for nonlinear Volterra systems with infinite impulse response. Given noisy fragmented input-output data, we identify the ``simplest" nonlinear model which best describes the data. The performance of the proposed approach is illustrated in two academic examples.

We believe the proposed method is useful in parsimonious Volterra system identification and control problems.  
In cases which we do not have much prior knowledge of system rather than the noisy collected data, this method provides an efficient way to estimate a model. 
Effort is now being put in the development of efficient algorithms that can both handle large amounts of data and  do not require the large number of variables that occur when using griding.

%

%
\bibliographystyle{plain}
\bibliography{ref10}
%
%

\end{document}